\documentclass{article}
\usepackage{spconf,amsmath,epsfig}

\title{CONDITIONAL GENERATION OF CLOUD FIELDS}
\twoauthors
 {Naser Mahfouz, Yi Ming}
	{Princeton University\\
	Atmospheric and Oceanic Sciences\\
	NOAA Geophys. Fluid Dynamics Lab. \\
	Division of Atmospheric Physics\\
	Princeton, NJ, USA}
 {Kaleb Smith}
	{NVIDIA Corporation\\
	Higher Education Research\\
	USA}
\begin{document}
%
\maketitle
%

\begin{abstract}
    Processes related to cloud physics constitute the largest remaining scientific uncertainty in climate models and projections.
    This uncertainty stems from the coarse nature of current climate models and relatedly the lack of understanding of detailed physics.
    We train a generative adversarial network to generate realistic cloud fields conditioned on meterological reanalysis data for both climate model outputs as well as satellite imagery.
    While our network is able to generate realistic cloud fields, especially their large-scale patterns, more work is needed to refine its accuracy to resolve finer textural details of cloud masses to improve its predictions. 
\end{abstract}
\section{Introduction}\label{sec:intro}

Clouds play an important role in regulating the Earth's radiation budget, and so our climate.
Indeed, the most significant portion of the remaining scientific uncertainty in climate models and projections is related to aerosol--cloud--climate interactions~\cite{seinfeld2016improving, stocker2014climate}.
Said uncertainty stems from the coarse resolution of (computationally practical and feasible) climate models which contributes to the lack of definitive understanding of aerosol--cloud processes spanning multiple scales.
Climate models are usually on the order of tens or hundreds of kilometers in resolution, whereas aerosol--cloud interactions typically manifest on the orders of micrometers, \(\mathcal{O}10^{-6}\) m, all the way to continental scales, \(\mathcal{O}10^{6}\) m~\cite{wood2012stratocumulus}.

Of particular interest are low-lying cloud fields as there is no clear conclusion how they will respond to a warming climate~\cite{brient2016shallowness, schneider2019possible, mulmenstadt2021underestimated}. Essentially, it is unclear if low-lying cloud fields will increase or decrease in response to the ongoing climate change, that is, further escalating the warming or blunting it.
As such, constraining and better understanding cloud fields as well as aerosol--cloud interactions will have a profound impact in improving the fidelity of climate projections.

We follow and extend previous works using neural networks to map meteorological conditions to cloud fields.
We utilize generative adversarial networks (GANs) to generate realistic cloud fields from real meteorological conditions. 
Recently, GANs have been used to recreate the Moderate-Resolution Imaging Spectrometer (MODIS) reflectance fields conditioned on meterological data from the Modern-Era Retrospective analysis for Research and Applications, Version 2 (MERRA-2)~\cite{yuan2019artificial, schmidt2020modeling}. 
Additionally, GANs have been used to recreate two-dimensional CloudSat vertical structures conditioned on MODIS and meterological conditions~\cite{leinonen2019reconstruction}.

Unlike previous works, we utilize model data produced by the fourth-generation atmosphere model at the National Oceanic and Atmospheric Administration's Geophysical Fluid Dynamics Laboratory (GFDL-AM4)~\cite{zhao2018gfdl1, zhao2018gfdl2}.
We find that we are able to use GANs to reproduce low-cloud fields in the Pacific Ocean and globally, as conditioned on meterological conditions from the NCEP/NCAR collection of reanalysis data~\cite{kalnay1996ncep}.
We then use the same set of conditions to probe the ability of GANs to recreate cloud fields in the Pacific Ocean as observed by the Geostationary Operational Environmental Satellite (GOES-17)~\cite{mccorkel2019goes}.

\section{Methodology}\label{sec:method}

Generative adversarial networks (GANs) rely on establishing a minimax game between two neural networks which work in opposition to each other~\cite{goodfellow2014generative}.
One network (often called a generator or creator) is trained to produce synthetic data whereas the opposing network (often called discriminator or critic) is trained to judge the authenticity (that is, real or generated) of the created data.
Optimally, the minimax game ends when the discriminator network's performance is equivalent to random guessing: The probability of its success is 50\%. 
A conditional GAN (CGAN) is a GAN whose generator and discriminator are provided with a set of conditions to create or judge the created data, respectively~\cite{mirza2014conditional}.

Mathematically, a generator \(\mathcal{G}\) has noise \(z\) and conditions \(c\) as inputs; and a discriminator \(\mathcal{D}\) has the generated data \(\mathcal{G}(z|c)\) or real data \(y\) as well as conditions \(c\) as inputs.
Then, the objective of the minimax game could be formulated with a gradient penalty on the Wasserstein distance-base formulation~\cite{arjovsky2017wasserstein} as in WGAN-GP~\cite{gulrajani2017improved} for efficient training 
\begin{equation}\nonumber
    \min_{\mathcal{G}} \max_{\mathcal{D}} \mathbf{E}(\mathcal{D}(\mathcal{G}(z|c)|c)) - \mathbf{E}(\mathcal{D}(y|c)) + \omega \mathbf{E}(\nabla \mathcal{D}(\mathcal{G}(z|c)))
\end{equation}
where \(\mathbf{E}\) is the notation for the expected value of the judgement produced by \(\mathcal{D} \) and \(\omega \) is a weighting factor for the gradient penalty.

We alternately train the generator and discriminator networks. 
For the generator, we use a U-Net-like design~\cite{ronneberger2015u, zhou2018unet++}, where on one side we draw random noise \(z\) and process the standardized conditions \(c\), propagating them together through convolution and deconvolution layers to produce the cloud fields \(\mathcal{G}(z|c)\).
For the discriminator, as is customary, we simply use a convolutional neural network until reaching a single-valued judgement \(\mathcal{D} \).
We note that \(\mathcal{D} (a|b) \) could be defined as a ``probability'' (that is, \( 0 \leq \mathcal{D} (a|b) \leq 1 \)) of \(a\) given \(b\), but for better training performance, we follow WGAN-GP as defined above, where its range is extended  \( -\infty \leq \mathcal{D} (a|b) \leq +\infty \).
We further modify the loss formulation above by adding a regularizing loss (mean squared error) on top of both the generator's and discriminator's losses, but with opposite signs.
 
\section{Results}\label{sec:results}

\subsection{GFDL-AM4 data}\label{sec:gfdl}

\begin{figure}[htb]
    \begin{minipage}[b]{1.0\linewidth}
      \centering
        \centerline{\epsfig{figure=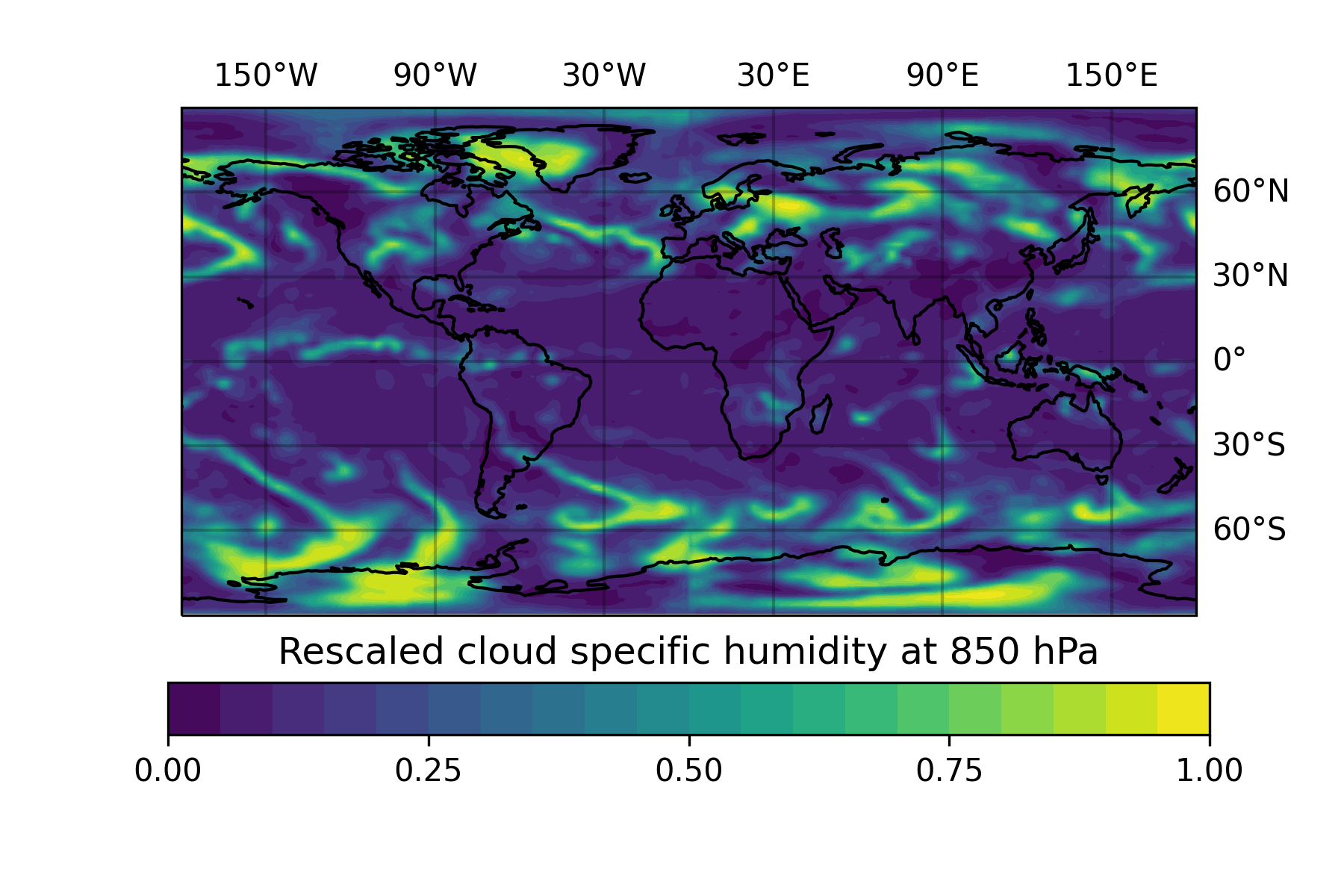,width=8.5cm}}
        \centerline{\epsfig{figure=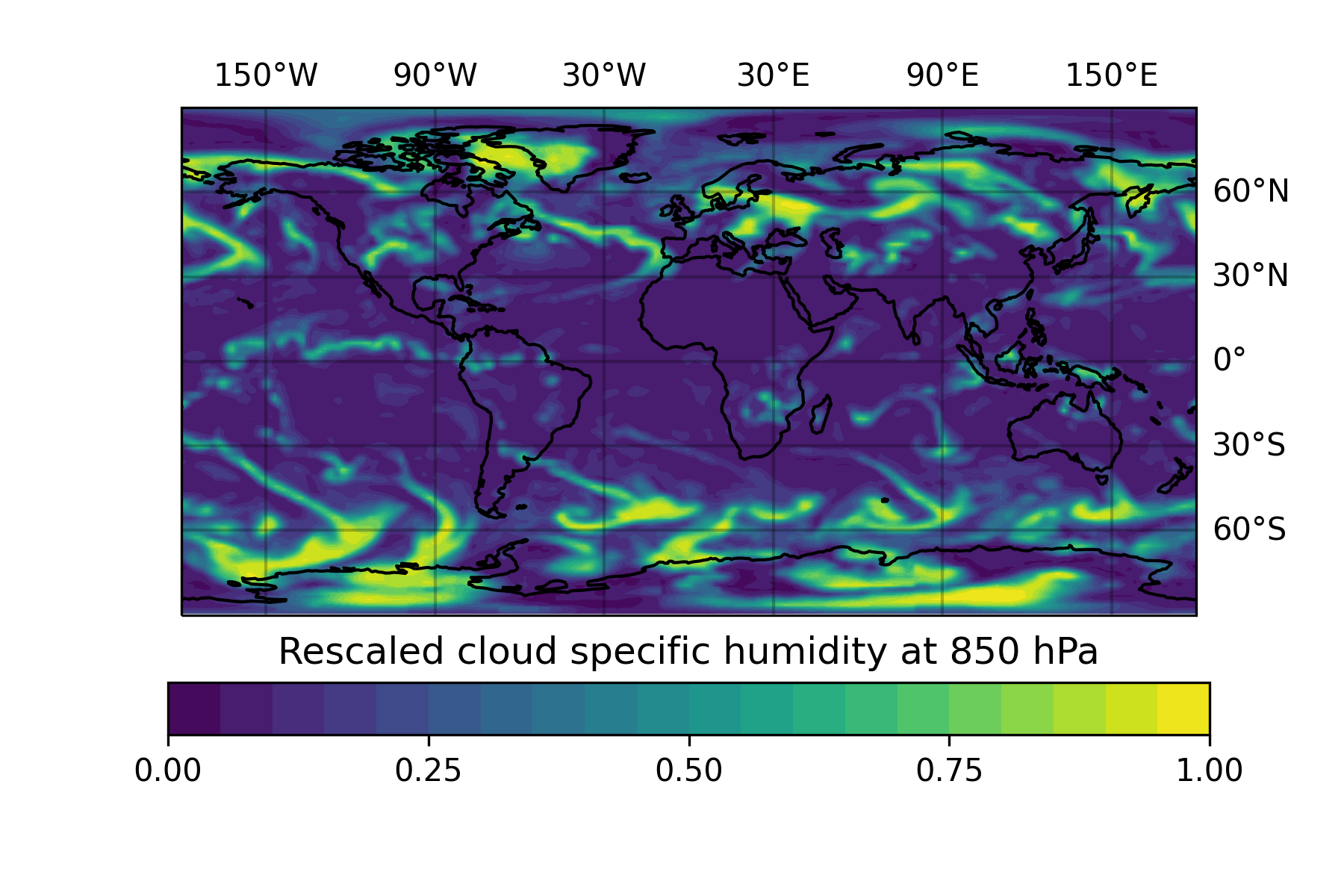,width=8.5cm}}
    \end{minipage}
    \caption{The CGAN-predicted outputs based on GFDL-AM4 data are in the top panels and the ground truths in the bottom panels. We find the trained CGAN is able to robustly recreate the cloud fields in this region for both the training and validation datasets (the validation dataset is never used during training of the neural networks).}\label{fig:gfdl}
\end{figure}

We use GFDL-AM4 to conduct a global climate nudged simulation~\cite{ming2021assessing} for the year 2010 and we use the corresponding NCEP/NCAR meteorological data as inputs to our CGAN.\@
For now, we elect to use the wind profile and temperature data from the NCEP/NCAR collection.
Both the nudged GFDL-AM4 simulation output as well as the NCEP/NCAR data used are provided four times daily at 0000, 0600, 1200, and 1800 UTC.\@
From the GFDL-AM4 output, we use the cloud liquid specific humidity at 850 hPa (kg/kg) as proxy for cloud cover, which we plot in Fig.~\ref{fig:gfdl}.
We find our prototypal CGAN to be robust enough, however more textural refinement is desired especially in the finer details of cloud masses in the global scale.
Moving from a limited regional domain to the global domain, we find that we need to increase the training times significantly to more accurately recreate the cloud scenes; in Fig.~\ref{fig:gfdl}, the results are plotted after 3500 full iterations (epochs) over a dataset containing 1314 scene--conditions pairs for training.

\subsection{GOES-17 data}\label{sec:goes}

\begin{figure}[htb]
    \begin{minipage}[b]{1\linewidth}
        \centering
            \centerline{\epsfig{figure=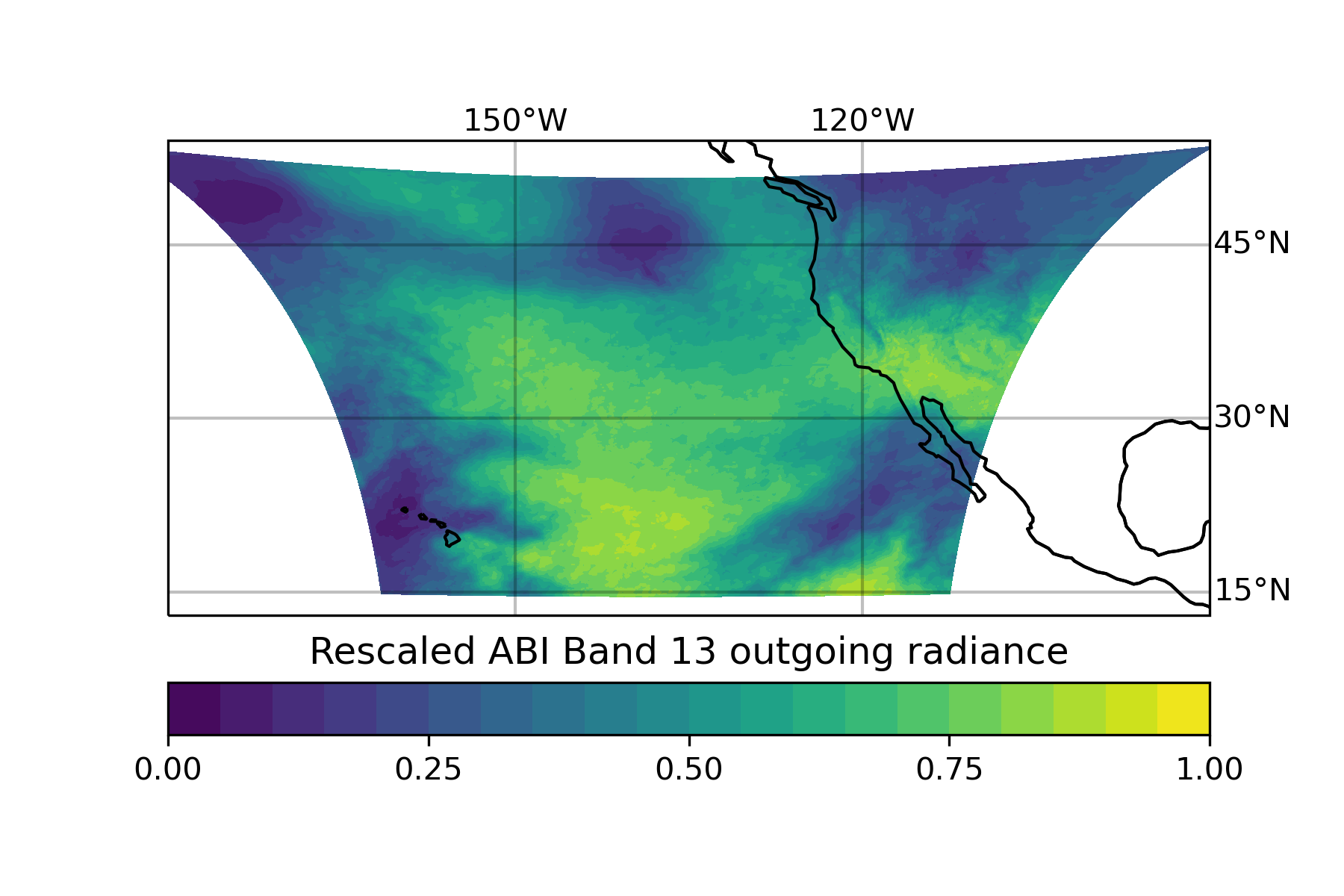,width=8.5cm}}
            \centerline{\epsfig{figure=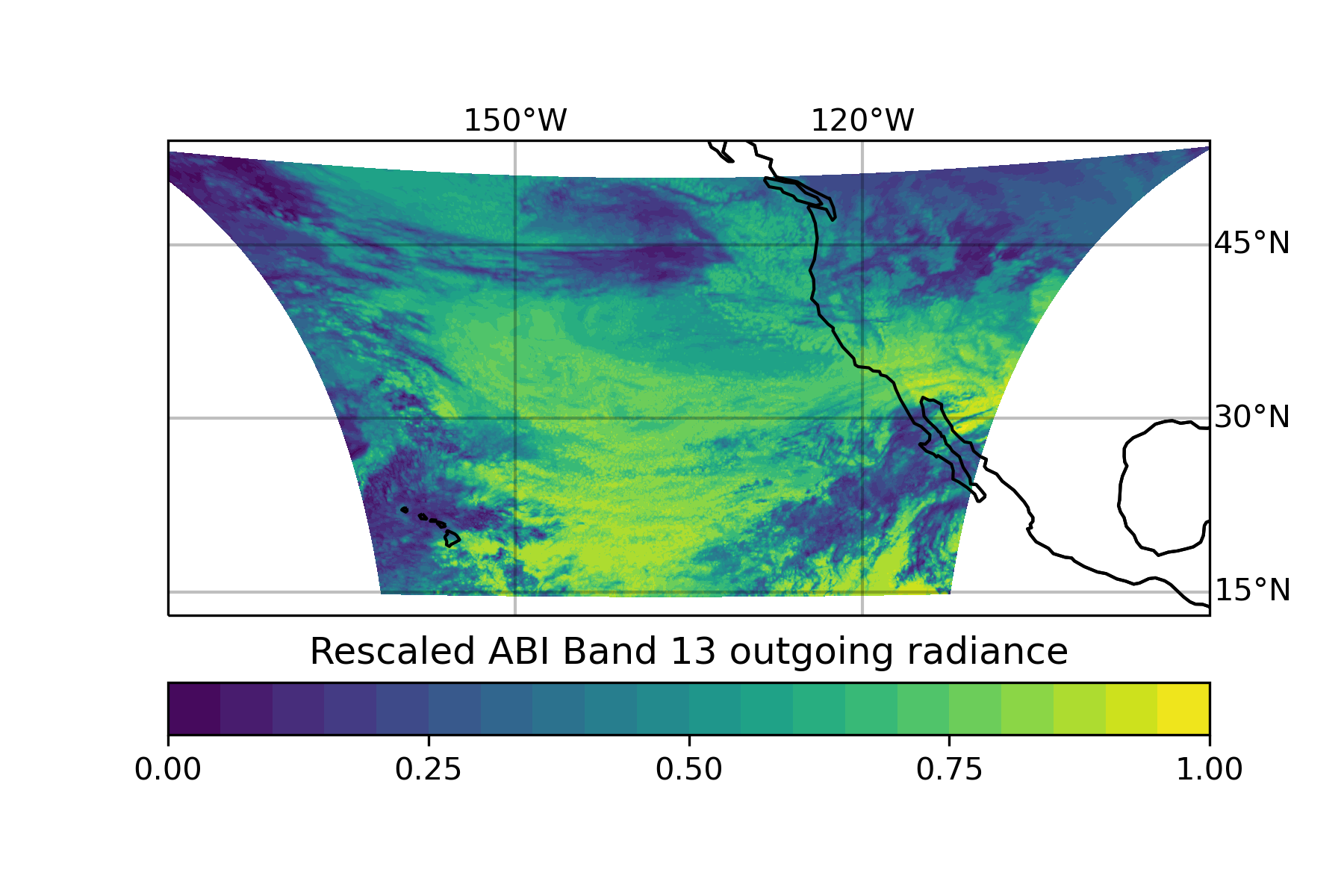,width=8.5cm}}
    \end{minipage}
    \caption{The CGAN-predicted output based on GOES-17 data are in in top panels and the ground truth are in the bottom panels for the ``CONUS'' imagery (Pacific Ocean). The trained CGAN is able to reproduce significant portion of the cloud fields as well as land masses (continental Americas as well as Hawaii) but its performance degrades when encountering finer textural details like those near in the middle.}\label{fig:goes}
\end{figure}

We use the same set of meteorological conditions (wind profile and temperature) from NCEP/NCAR collection of reanalysis data to train GANs for the Pacific Ocean.\@
However, this time around, we use the NCEP/NCAR data from 2020 alongside GOES-17 ``CONUS'' imagery (over the Pacific Ocean) at 0000, 0600, 1200, and 1800 UTC.\@
We utilize the ``clean'' IR longwave window (Advanced Baseline Imager Band 13) that is able to detect to detect and track clouds throughout the day~\cite{schmit2017closer}.
We retrain the CGAN model with the same architecture, except with a new dataset comprising 15011 scene--condition pairs.\@
As shown in Fig.~\ref{fig:goes}, we find that the retrained CGAN can recreate the cloud fields in the Pacific Ocean, albeit with a lower quality than in Section~\ref{sec:gfdl}.
In particular, we note that the retrained CGAN can generally recreate the large-scale patterns of cloud masses as well as land (for example the western coast of the USA and even Hawaii; here shown with the outlines of the coastlines, but it can be seen clearly otherwise).
However, its performance suffers in resolving finer details like those near the middle.
Training the CGAN on satellite data is significantly costlier than training it on GFDL-AM4 data; in Fig.~\ref{fig:goes}, the results are produced after only 700 full iterations (epochs) over a dataset containing 15011 scene--conditions pairs for training.
For comparison, the GFDL-AM4 data resolution is 188\(\times \)288 (resized inside the network to 128\(\times \)256) whereas the resolution of the satellite data is 1500\(\times \)2500 (resized to 256\(\times \)512).
Due to computational graph optimizations in our networks, the resizing is based on the simple ``nearest neighbor'' interpolation for now.

\section{Discussion}\label{sec:discussion}

In this work, we use a conditional generative adversarial network (CGAN) to generate realistic cloud fields based on meterological conditions from reanalysis data.
We rely on reanalysis meterological data as conditions; and we use nudged climate model simulations and satellite imagery as inputs in training.
Our CGAN prototype thus far can reproduce large portions of the cloud fields, especially large-scale patterns of cloud masses.
However, it is not yet able to reproduce the finer textural details of these cloud fields, especially when applied to high-resolution satellite data.
Our approach is distinct from previous works in that it relies on climate model outputs as well as the unedited, L1b, outputs of new generation of geostationary satellites. 

For the rest of this section, we discuss our current and future work, as well as potential implications and challenges ahead.
We provide the full code for the CGAN model architecture and a sample dataset for reproducibility~\cite{naser_mahfouz_2022_6581540}.
We thank Pu Lin for the GFDL-AM4 simulation data and the GFDL Modeling Systems Division for access to compute resources.
 
\subsection{More channels and conditions}

For now, we use a single channel for both the GFDL-AM4 data in Section~\ref{sec:gfdl} the satellite from GOES-17 in Section~\ref{sec:goes}.
However, in order to improve the accuracy of CGAN --- especially when it comes to finer textural details --- we can use more channels.
For example, we could utilize another IR channel (for example, shortwave, 13) from the GOES-17 imagery as well as the RGB ones; likewise, we can use other parameters from the climate model output data such as cloud cover and liquid water path.
The same applies to reanalysis data; for now, we elect to use only the wind profile and temperature, but we can extend our election to other parameters that are relevant, including their vertical distributions.

\subsection{Domain expertise and transfer learning}\label{sec:transfer}

We utilize domain expertise in selecting the conditions from the reanalysis data; we select wind profile and surface temperature because we think they will likely determine cloud fields.
It is possible to extend the application of domain expertise in this problem, for example, by feeding the CGAN with predetermined cloud patterns that are known to be present and classified in the field.
This can, in theory, enable us to use transfer learning and style transfer to improve the accuracy and training of CGAN~\cite{zhuang2020comprehensive}.

\subsection{Resolutions and scales}

This problem represents a challenge across resolutions and scales.
Clouds start growing around submicron particles and then become larger and larger to span continental scales.
Therefore, changes on the submicron scale can ultimately effect changes on the continental scale.
Additionally, reanalysis data is only provided with a relatively coarse resolution (even compared with coarse modern climate models) and thus the CGAN model must be trained to convert the coarsely resolved conditions alongside and toward the more finely resolved resolution of climate model and satellite data.
We postulate that implementing innovative elements of other generative modeling techniques such as transfer learning as discussed briefly in Section~\ref{sec:transfer}, super-resolution~\cite{ledig2017photo}, down-scaling~\cite{leinonen2020stochastic}, and now-casting~\cite{ravuri2021skillful} can help in ultimately improving the accuracy and fidelity of the CGAN used herein.

\vfill
\pagebreak

\bibliographystyle{IEEEbib}
\bibliography{strings,refs}

\end{document}